\documentclass[12pt]{iopart}
\usepackage{mathptmx}
\usepackage{graphicx}
\usepackage[table]{xcolor}
\definecolor{darkred}{rgb}{0.5,0,0}
\definecolor{darkblue}{rgb}{0,0,0.5}
\definecolor{firebrick}{rgb}{0.75,0.125,0.125}
\definecolor{darkgreen}{rgb}{0,0.5,0}
\usepackage[colorlinks=true,linkcolor=firebrick,citecolor=darkgreen,urlcolor=darkblue]{hyperref}
\usepackage[font=footnotesize]{caption}
\usepackage[left=1.5cm,right=1.5cm,top=2.5cm,bottom=2.5cm]{geometry}
\expandafter\let\csname equation*\endcsname=\relax 
\expandafter\let\csname endequation*\endcsname=\relax 
\usepackage{amsmath,amssymb}
\usepackage{orcidlink}
\usepackage{upgreek}
\usepackage{lineno}

\def\avg#1{{\langle{#1}\rangle}}

\begin{document}

\title[Calculation of rescaling factors and nuclear
multiplication of muons \ldots]{Calculation of rescaling factors and nuclear
multiplication of muons in extensive air showers}

\author{
Kevin Almeida Cheminant$^1$\orcidlink{0000-0001-6352-5339},
Dariusz G\'ora$^1$\orcidlink{0000-0002-4853-5974},
Nataliia Borodai$^1$\orcidlink{0000-0003-1864-937X},
Ralph Engel$^2$\orcidlink{0000-0003-2924-8889},
Tanguy Pierog$^2$\orcidlink{0000-0002-7472-8710},
Jan P\c{e}kala$^1$\orcidlink{0000-0002-1062-5595},
Markus Roth$^2$\orcidlink{0000-0003-1281-4477},
Michael Unger$^2$\orcidlink{0000-0002-7651-0272},
Darko Veberi\v{c}$^2$\orcidlink{0000-0003-2683-1526} and
Henryk Wilczy\'nski$^1$\orcidlink{0000-0003-2652-9685}
}

\address{$^1$Institute of Nuclear Physics PAS \raisebox{-0.5ex}{\includegraphics[height=2.3ex]{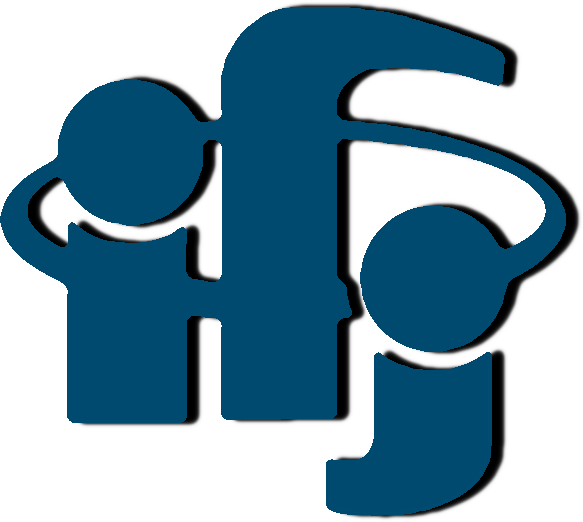}}, Radzikowskiego 152, Krakow, Poland
\\
$^2$Karlsruhe Institute of Technology \includegraphics[height=1.55ex]{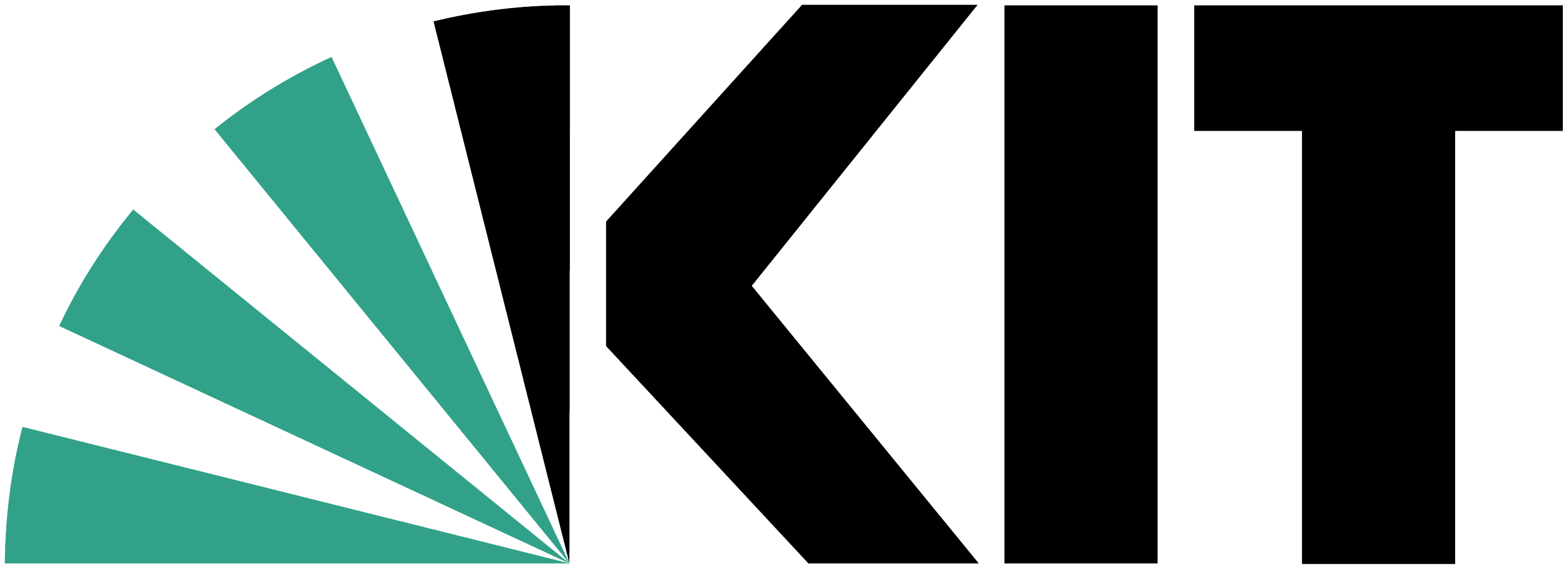}, Institute for Astroparticle Physics, Karlsruhe, Germany}
\ead{\href{mailto:dariusz.gora@ifj.edu.pl}{dariusz.gora@ifj.edu.pl}}
\vspace{10pt}
\begin{indented}
\item[]September 2022
\end{indented}

\begin{abstract}
Recent results obtained from leading cosmic ray experiments indicate that simulations using LHC-tuned hadronic interaction models underestimate the number of muons in extensive air showers compared to experimental data. This is the so-called muon deficit problem. Determination of the muon component in the air shower is crucial for inferring the mass of the primary particle, which is a key ingredient in the efforts to pinpoint the sources of ultra-high energy cosmic rays. In this paper, we present a new method to derive the muon signal in detectors, which uses the difference between the total reconstructed (``data'') and simulated signals, and is  in turn related to the muon signal which  is roughly independent of the zenith angle, but depends on the mass of the primary cosmic ray. Such a method offers an opportunity not only to test/calibrate the hadronic interaction models, but also to derive the $\beta$ exponent, which describes an increase of the number of muons in a shower as a function of the energy and mass of the primary cosmic ray. Detailed simulations show a dependence of the $\beta$ exponent on hadronic interaction properties, thus the determination of this parameter is important for understanding the muon deficit problem.
We validate the method by using Monte-Carlo simulations for the EPOS-LHC and QGSJetII-04 hadronic interaction models, and  show that this method allows us to recover the ratio of the muon signal between EPOS-LHC and QGSJetII-04 and the average $\beta$ exponent for the studied system, within less than a few percent. This is a consequence of the good recovery of the muon signal for each primary included in the analysis.
\end{abstract}

%
%
%
%
%

\section{Introduction}

Discovered at the beginning of the 20th century by  Victor F.~Hess,
cosmic rays are protons and atomic nuclei that constantly bombard Earth's atmosphere. Before arriving to the surface of Earth they first interact with the  nuclei of the atmosphere to produce cascades of secondary particles that may develop all
the way to the ground. This physical phenomenon, also called extensive air shower
(EAS), can be detected via multiple channels of observations, e.g.\ Cherenkov and 
fluorescence light, or radio emission, which can measure different physical quantities
that can be used to determine the nature of the primary cosmic ray, its arrival direction, and its energy. 

In the initial phase of the cascading process, the number of particles increases while the energy per particle drops and distinct components emerge, namely the hadronic, electromagnetic, and
muonic components. Such growth carries on until a maximum is reached,  at a traversed depth usually referred to as $X_\text{max}$, as particles below a certain energy threshold
are no longer capable of producing additional particles  but decay instead as atmospheric absorption
processes start taking over. As many as $10^6$  to $10^9$ secondary
particles may reach the ground over an area that can extend up to several square kilometers.

\begin{figure}
  \centering
  \includegraphics[width=0.8\linewidth]{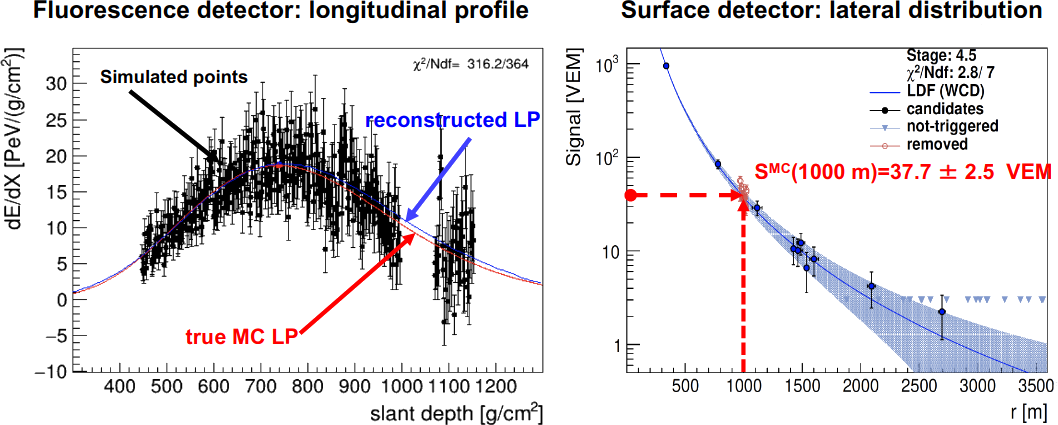}
  \caption{Example of TD -- simulated event. \textbf{Left}: The measured longitudinal profile  (LP) of an illustrative air shower with the matching simulated showers, using QGSJetII-04 for proton  with the energy $10^{19}$\,eV.  The longitudinal profile for true simulated event (true MC LP) and  reconstructed LP are  also shown.  
  \textbf{Right}: The simulated ground signals for the same event. The signal at 1000\,m ($S_{1000}$) in VEM is also shown. 1\,VEM corresponds to the most-likely signal deposited by a muon that traverses vertically the center of the SD station. }
  \label{fig:fitauger}
\end{figure}

In order to describe how EAS are formed in the atmosphere, simple toy models,
such as the one described by Heitler and Matthews~\cite{heitler}, have been developed and are
capable of providing accurate predictions of some of the  main quantities that characterize
air showers without the need for high-performance computing. Although simplistic, the Heitler-Matthews model is powerful enough to allow the discrimination of EAS produced by protons/nuclei and photons.  The number of muons $N_\upmu^A$ in EAS  initiated by a nucleus with mass
number $A$ can be related to the number of muons $N_\upmu^\text{p}$ produced in a shower initiated by a proton with the same energy through $N_\upmu^A=N_\upmu^\text{p}\, A^{1-\beta}$, where $1-\beta\simeq 0.1$.  Muons in EAS have also large decay lengths and small radiative energy losses and are produced at different stages of  the shower development. Therefore, muons can reach surface and underground detector arrays while keeping relevant information about the hadronic cascade.

In recent years, our understanding of the nature of cosmic rays has 
significantly improved thanks to experiments spread out all over the world and using different methods of detections such as gamma-ray telescopes (H.E.S.S.~\cite{hess}, MAGIC~\cite{magic},VERITAS~\cite{veritas}, HAWC~\cite{hawc}, and others) and cosmic-ray observatories (Telescope Array~\cite{ta2}, Pierre Auger Observatory~\cite{auger}). As of today, the energy spectrum of cosmic rays has been measured from a few GeV (giga electron-Volts, $10^9$\,eV) up to 100\,EeV (100 exa electron-Volts, $10^{20}$\,eV), well beyond the energy accessible in terrestrial particle accelerators, and falls rapidly as the energy increases. The general consensus is that cosmic rays below $10^{17}$ to $10^{18}$\,eV are of Galactic origin, most likely from supernovae, while particles above this energy range have their origin in extra-galactic sources, with active galactic nuclei and starburst galaxies being the most plausible candidates.

Simulations of EAS using current hadronic interaction models predict fewer  muons than observed  in  real events, which is known as the muon deficit problem~\cite{denis}. As an example, data from the Pierre Auger Observatory
indicate that the muon number predicted by the LHC-tuned  models, such as EPOS-LHC~\cite{epos} and QGSJetII-04~\cite{qgsjet}, is 30\% to 60\% lower than what is observed in showers with an energy of $10^{19}$\,eV~\cite{allen}. The muon excess over predictions seen  by the Pierre Auger Collaboration is/was also seen in  several other experiments like HiRes/MIA~\cite{hires}, NEVOD-DECOR~\cite{nevod}, SUGAR array~\cite{sugar}, Telescope Array~\cite{ta}. However,   experiments like KASCADE-Grande~\cite{cascade} and EAS-MSU~\cite{EASMU} reported  no discrepancy in the muon number around $10^{17}$\,eV. In Ref.~\cite{hans},  after cross-calibration of the energy scales,
the observed muon densities were scaled by using the so-called $z$-scale and compared to expectations from different hadronic models, also for  data from IceCube~\cite{icecube} and AMIGA~\cite{amiga}.  While such densities were found to be consistent with simulations up to $10^{16}$\,eV, at higher energies the muon deficit increases in several experiments~\cite{hans}. Since data interpretation relies on simulations, the muon deficit problem has deep implications: the data suggest a much heavier composition of cosmic rays based on muons only than the composition derived from $X_\text{max}$ measurements~\cite{xmax}. 

To study the muon-number problem, a top-down (TD) reconstruction method was proposed by the Pierre Auger Collaboration~\cite{allen} for the so-called hybrid events (events seen simultaneously by the array of particle detectors (SD)  and by the fluorescence detector (FD)).
The main aim of the TD reconstruction is to predict signals in the  FD and SD on a simulation basis. In the TD method, one finds a simulated shower, which has a  distribution of electromagnetic component along the shower axis (longitudinal profile,  reconstructed LP) most similar to the observed longitudinal profile of the shower (i.e.\ reference profile (true MC LP) -- see left panel of Fig.~\ref{fig:fitauger}). The reference longitudinal profile is linked to the electromagnetic component of the shower, so the method relies on the fact that this component is accurately simulated. 

As an output, the TD method provides a reconstructed event, in which the signal in the SD is determined using Monte-Carlo (MC) simulations. The simulated SD signals in the output shower, which depend on the interaction model, may then be compared with the data/initial shower. The SD signal includes the contribution of muons, which are tracers of properties of the hadronic interactions. A comparison of the simulated SD signal with the corresponding signal in the data shower provides an opportunity to check the correctness of the lateral distribution of the simulated showers (see right panel of Fig.~\ref{fig:fitauger}). Since the lateral distribution is sensitive to the hadronic interaction models used, an analysis of this distribution provides an opportunity to investigate indirectly the interaction models at energies much  above the maximum energies provided by  terrestrial accelerators. It is therefore expected that the TD method should allow us to calibrate the interaction models, and to reduce the discrepancy between the data and simulations.

The TD analysis performed in this work is similar to the one presented in  Refs.~\cite{tiff,goraicrc2021} and is based on the analysis found in Ref.~\cite{allen}.  In this work the TD chain includes a simulation of the  SD response for the CORSIKA~\cite{corsica} simulated
event\footnote{We use one of the latest version of CORSIKA, i.e.\ version 7560.} -- the reference shower. The Pierre Auger Observatory response for the reference shower is simulated in
the hybrid mode -- the event is seen by SD and FD -- using the Offline software~\cite{offline}  which provides 10  detector simulations for comparison of the station signals with the reference MC event. 

Here we also try to reproduce as accurately as possible the real data from the  Pierre Auger Observatory by creating a mock data set of mixed composition from MC simulations obtained with the EPOS-LHC hadronic model, at $10^{19}$\,eV. MC simulations at the same  energy produced using the QGSJETII-04 model are then used to try to recover the muon signal from this mock data set, by calculating the muon scaling factors  (relative to EPOS-LHC) for the primaries considered in this dataset. 

 In this paper, we present a validation test of the method for determining muon scaling factors by analysing reconstructions of a simulated hybrid shower, where mock-data showers are used as reference events. This method is based on the $z$-variable, which is  the difference between the initially simulated  and the reconstructed total signal at the detectors, 1000 meters away from the shower axis, and which is related to the muon signal. This variable is approximately independent of the zenith angle, but depends on the mass of the primary cosmic ray.  We show that we can recover the ratio of the muon signal between EPOS-LHC and QGSJetII-04, on average, within less than 6\%, and the average $\beta$ exponent which governs the number of muons in simulated air showers~\cite{heitler}, within less than 1\%, which is a consequence of the good recovery of the muon signal for each primary.

\section{Preparation of the mock dataset and QGSJetII-04 Monte-Carlo Simulations} \label{sec1}

\begin{figure}
  \centering
  \includegraphics[width=0.5\linewidth]{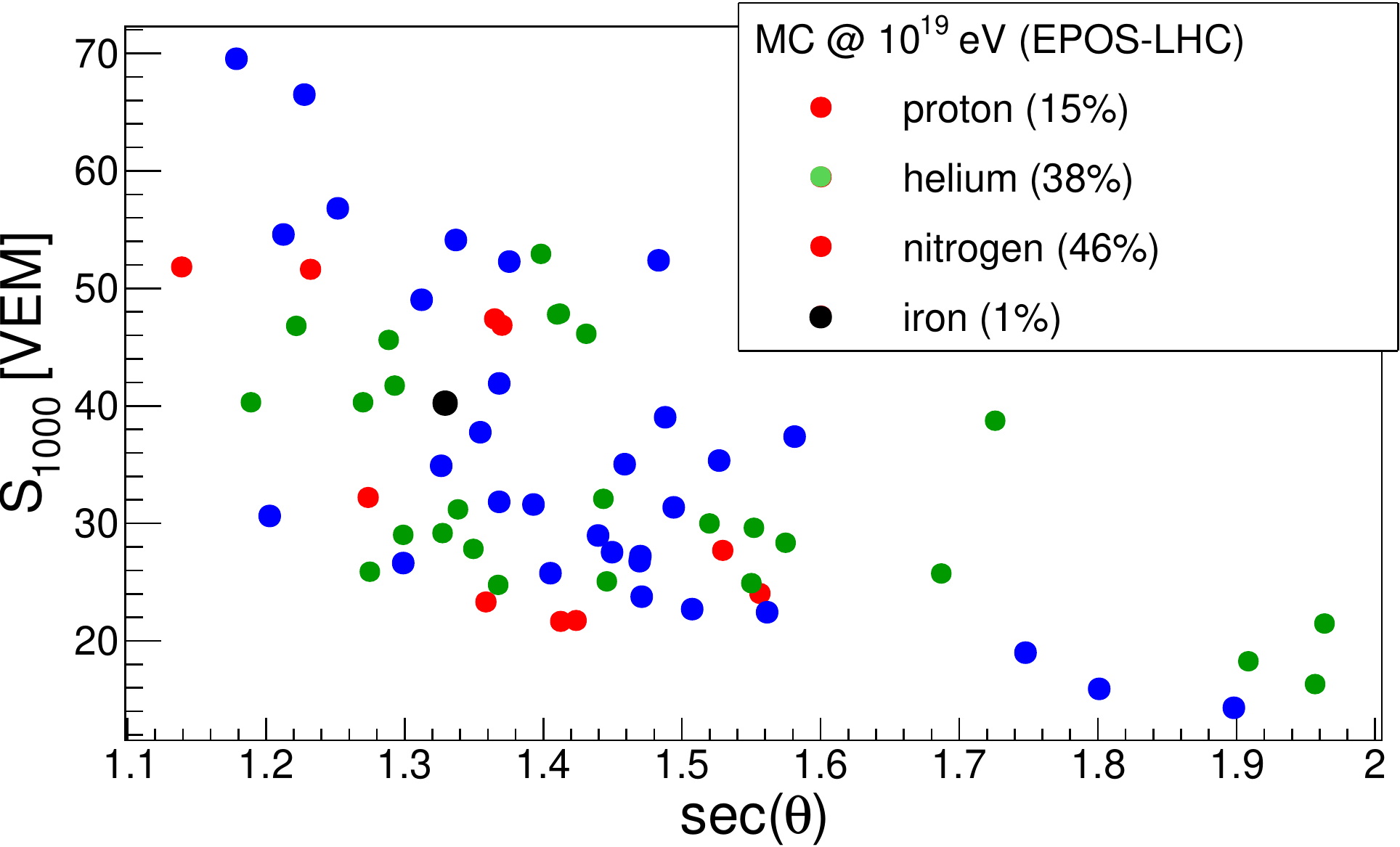}
  \caption{Total signal  $S_{1000}$ of the mock dataset shown with the proportion of primaries discussed in the text.}
  \label{fig:MOCK-DATA}
\end{figure}

The mock data set of EPOS-LHC is built based on simulated CORSIKA events. CORSIKA simulations are performed for four potential primaries: proton, helium, nitrogen, and iron.  If we restrict ourselves to shower energies $10^{18.8} < E < 10^{19.2}$\,eV to zenith angles $\theta$ below $60^\circ$, we get 68 reference events (typical number of high-energy, high-quality events observed over several years by the Pierre Auger Observatory in a narrow energy interval).  The total signal $S_{1000}$ for each  reference event as a function of $\sec\theta$  is shown in Fig.~\ref{fig:MOCK-DATA}.  As expected, due to the attenuation of EAS in the atmosphere, we see that the signal depends on the zenith angle.

To calculate the muon signal for each MC event, we use the universal parametrization of muon fraction in the signal $g_{\upmu,i}(\theta)$ presented in  Refs.~\cite{goraicrc2021,kegl}.  This parametrization gives the muon signal for a primary $i$ and a zenith angle $\theta$ such that $S_{\upmu,i}^\text{mock}(\theta)=g_{\upmu,i}(\theta) \, S_{1000,i}^\text{mock}(\theta)$. 

In order to create the mock dataset of mixed composition, we consider the fractions of primaries $f_i$ for proton, helium, nitrogen, and iron as measured by the Pierre Auger Observatory  at $10^{19}$\,eV from EPOS-LHC, as presented in Ref.~\cite{souza2016}. These fractions are roughly estimated to be around 15\%, 38\%, 46\%, and 1\% for proton, helium, nitrogen, and iron, respectively. 
In this work 68 reference CORSIKA events were selected from EPOS-LHC simulations, taking into account these primary fractions, which roughly translates to 10 proton events, 26 helium events, 31 nitrogen events, and 1 iron event. The average muon signal for the mock dataset is $17.30\pm0.25$, $19.03\pm0.30$, $21.12\pm0.38$, and $(23.42\pm0.25)$\,VEM for proton, helium, nitrogen, and iron primaries, respectively. 

\begin{figure}
  \centering
  \includegraphics[width=.47\linewidth]{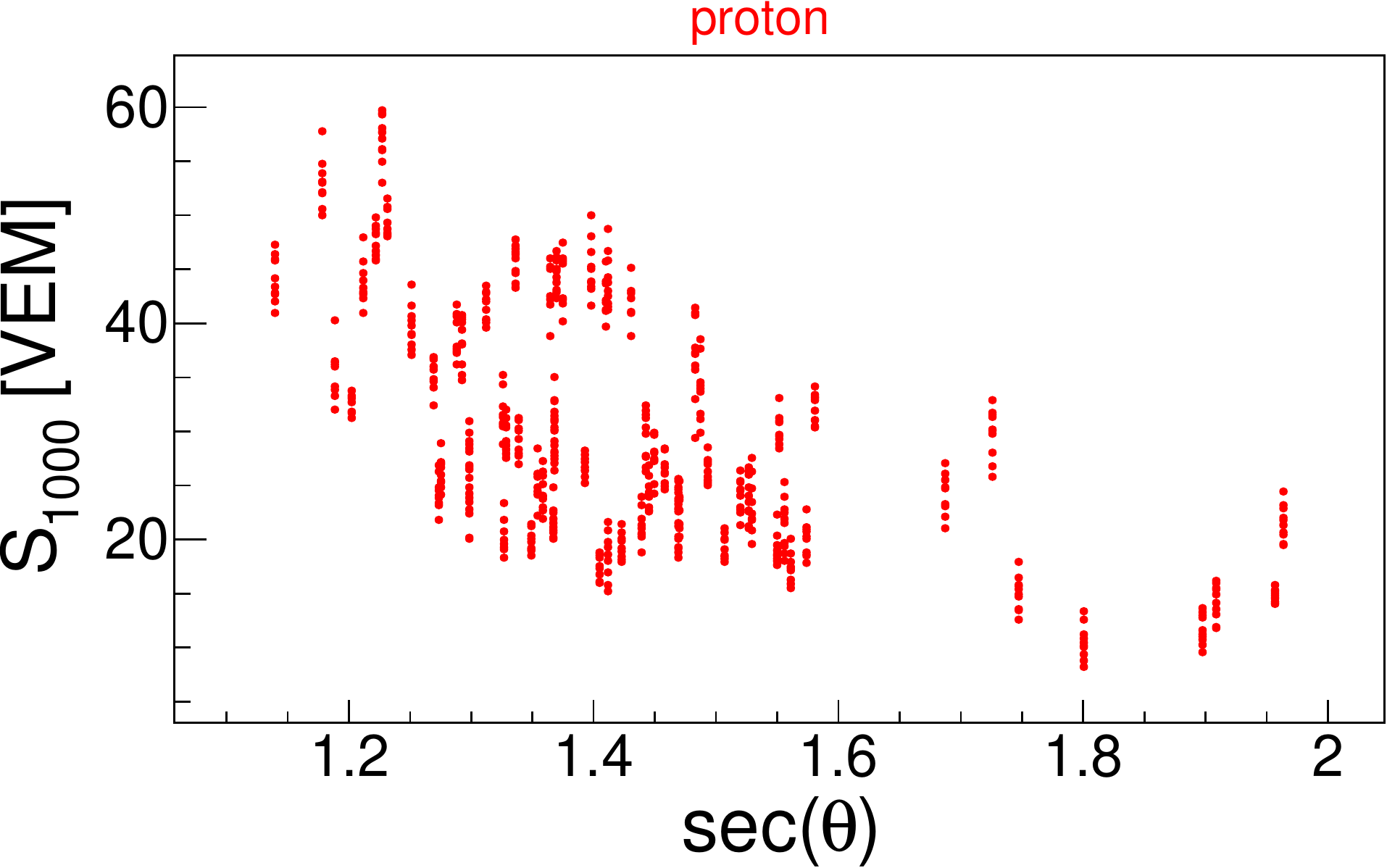}\hfill
  \includegraphics[width=0.48\linewidth]{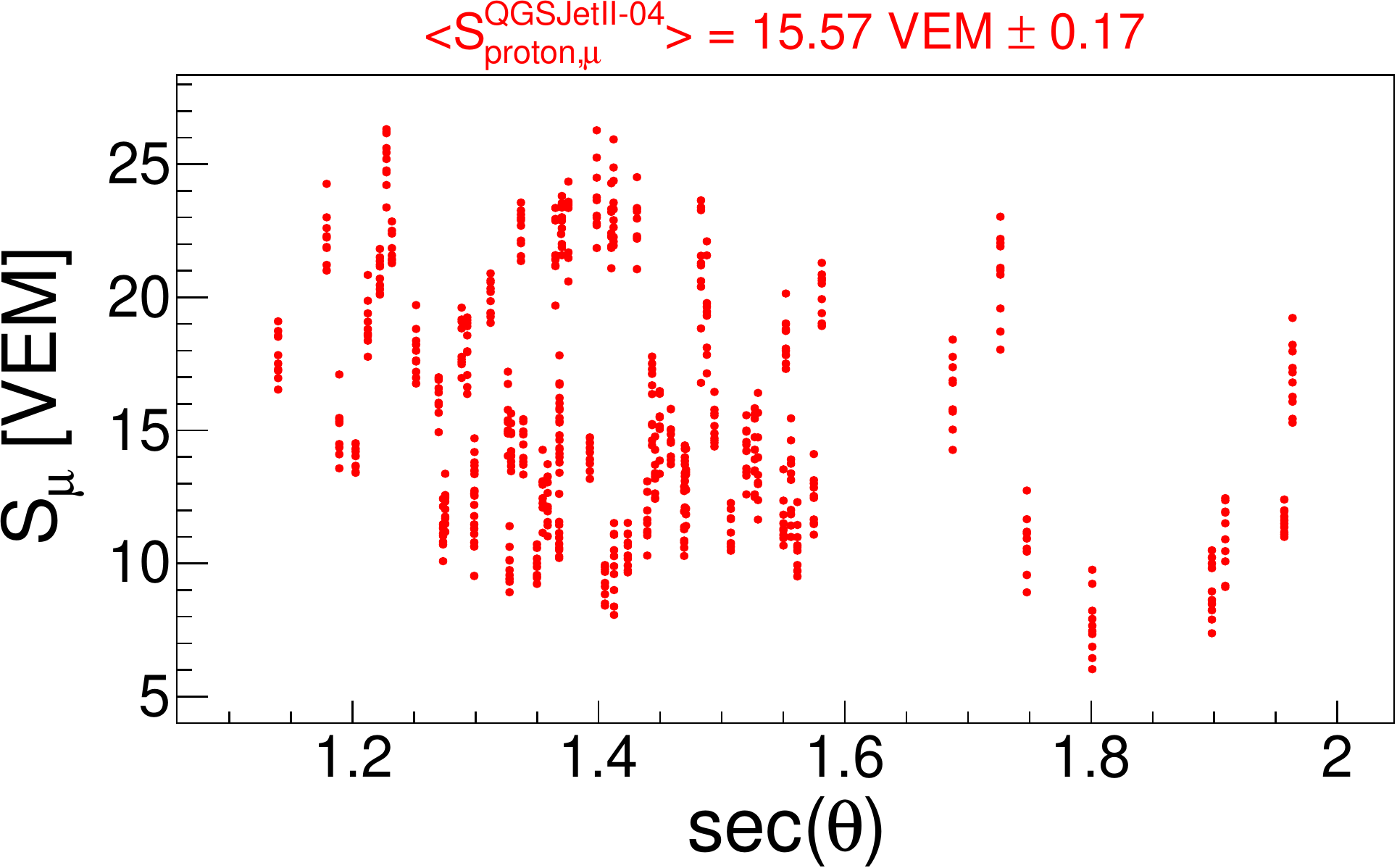}
  \caption{\textbf{Left}: Total signal $S_{1000}$ in events simulated with QGSJetII-04 for proton primary. The average total signal for proton is $(29.66\pm0.41)$\,VEM for the studied range of zenith angle, i.e.\ less than $60^\circ$. \textbf{Right:} Muon signal in events simulated with QGSJetII-04 for proton primary. The average muon signal for proton is $(15.45\pm0.17)$\,VEM.}
  \label{fig:4}
\end{figure}

In this TD analysis, we consider air shower simulations obtained with QGSJetII-04 model as our MC sample. It is used to reconstruct the muon signal found in the mock dataset described in the previous section. 10 MC showers are associated to each shower from the mock dataset with a given zenith angle, for each primary. As an example, the distributions of $S_{1000,ij}^\text{MC}$ and $S_{\upmu,ij}^\text{MC}=g_{\upmu,i}(\theta)\,S_{1000,ij}^\text{MC}$ as a function of $\sec\theta$ for QGSJetII-04 simulations of proton is shown in Fig.~\ref{fig:4}~(left) and Fig.~\ref{fig:4} (right), respectively. In this case, the average muon signal for each primary is $15.57\pm0.17$, $17.25\pm0.19$, $19.37\pm0.20$, and $(21.62\pm0.23)$\,VEM. One can observe that the average muon signal for each primary is larger for EPOS-LHC simulations than for QGSJetII-04. Thus  the mean ratio   averaged over the four primaries studied $r_\text{true}^\text{MC}=\avg{\avg{S_\upmu^\text{EPOS}}/\avg{S_\upmu^\text{QGS}}}$ is approximately equal to $1.10 \pm 0.04$. It is worth noting that  in this way, i.e.\ using EPOS-LHC as mock data set and MC simulation from QGSJetII-04, we also mimic the muon problem  seen in the real data. 

\begin{figure}
  \centering
  \def\h{0.29}
  \includegraphics[height=\h\linewidth]{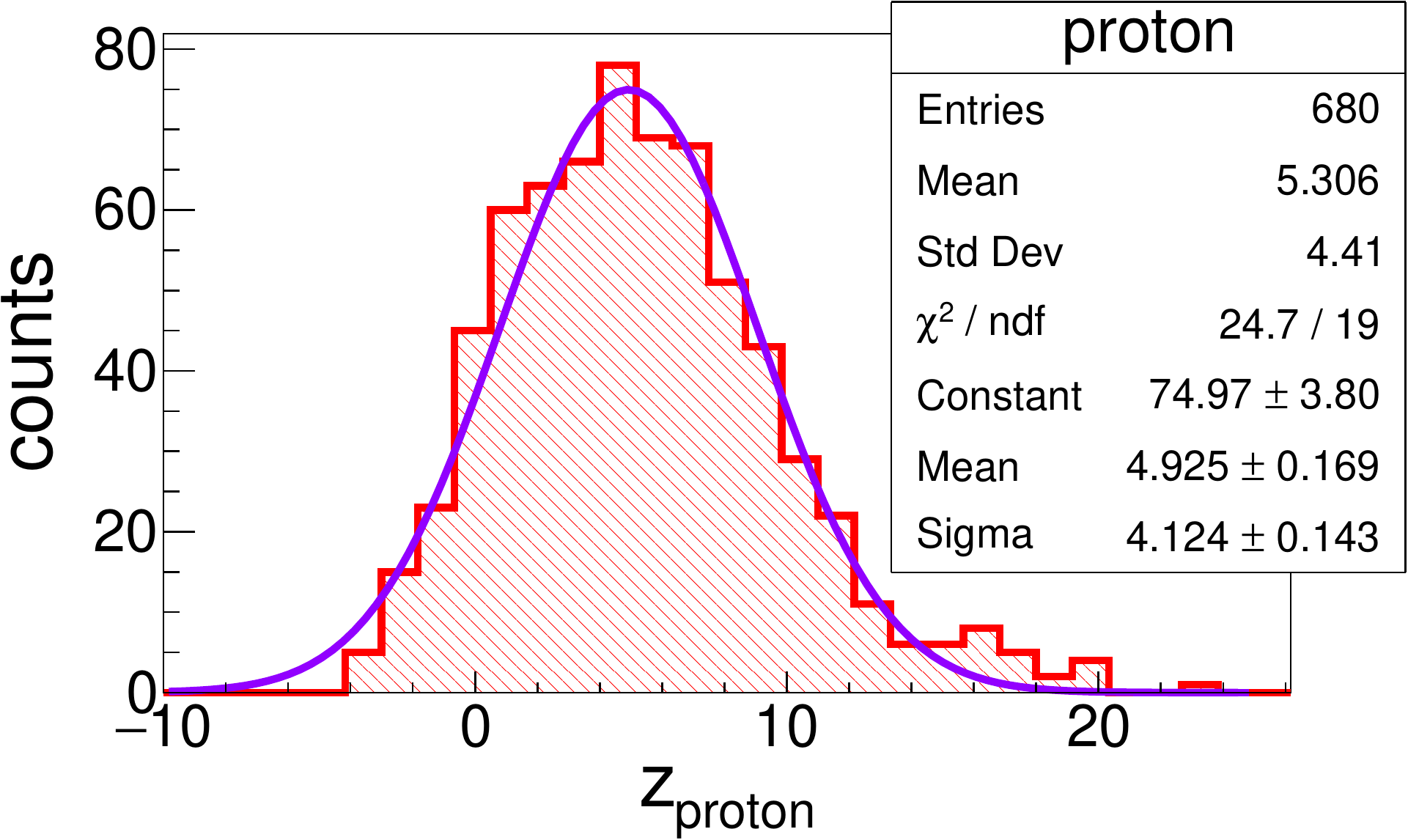}\hfill
  \includegraphics[height=\h\linewidth]{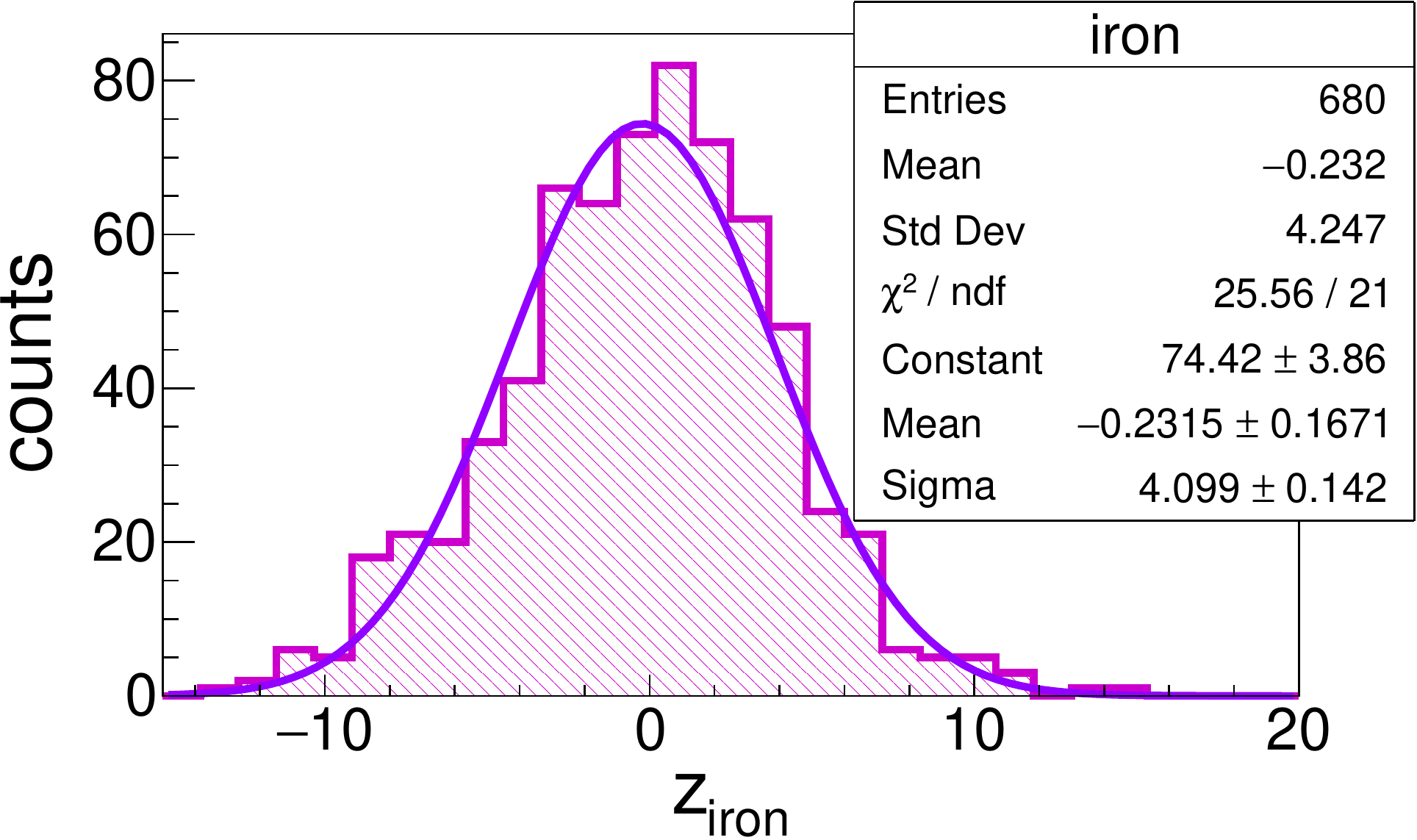}
  \caption{$z_i$ distributions for proton (left)  and iron primary (right)  based on mock data (EPOS-LHC) and QGSJetII-04 simulations.}
  \label{fig:44}
\end{figure}

\section{Muon scaling factor}

The observed SD signal of ultra-high energy air showers is significantly larger than predicted by hadronic models tuned to fit the accelerator data~\cite{allen}. Such a disagreement can be corrected for  by introducing linear scaling factors, for the electromagnetic part, $R_\text{EM}$, and the hadronic/muonic part, $R_\upmu$. Following this approach for a single shower $j$, the simulated ground signal at 1000\,m from QGSJetII-04 MC and the mock dataset can be written as
\begin{align}
S_{1000,j}^\text{MC} &\equiv
  S_{\text{EM},j}^\text{MC}+S_{\upmu,j}^\text{MC},
\label{s_1000}
\\
 S^\text{mock}_{1000,j}(R_\text{EM}, R_\upmu) &\equiv
   S_{\text{EM},j}^\text{mock} + S_{\upmu,j}^\text{mock} =
   R_\text{EM} \, S_{\text{EM},j}^\text{MC} + R_{\upmu,j}\,R_\text{EM}^\alpha \, S_{\upmu,j}^\text{MC}.
\label{s_mock}
\end{align}
In above Eq.~\eqref{s_mock} we have used the fact that some of the electromagnetic particles produced by muons in decay or radiation processes, as well as by low-energy $\uppi^0$s, can be attributed to the electromagnetic signal by introducing an additional factor $R_\text{EM}^\alpha$; but the muons that result from photoproduction are assigned to the electromagnetic signal, $S_\text{EM}$. As shown in Ref.~\cite{allen}, no rescaling is needed for the electromagnetic part, where the most likely solution is $R_\text{EM}=1$. Furthermore, in the TD method the reference longitudinal profile is related to the electromagnetic component of the shower, so the method ensures that this part is accurately simulated. Hence the assumption $R_\text{EM}=1$ used in this analysis. 

In this work, we use the difference between the mock dataset and the MC ground signal as the main observable, i.e.\ $z_j\equiv S_{1000,j}^\text{mock}-S_{1000,j}^\text{MC}$, because this variable is a natural indicator of the discrepancy between data and MC. Moreover, the discrepancy should ideally be zero. Another interesting feature arises from Eqs.~\eqref{s_1000} and \eqref{s_mock}. For $R_\text{EM}=1$,  by simple subtraction we obtain
\begin{equation}
S_{\upmu,ij}^\text{MC} = \frac{z_{ij}}{R_{\upmu,ij}-1},
\label{eqSmu}
\end{equation}
for a primary $i$ and an event $j$. This is the key equation for the method presented in this paper. The formula shows that the muon MC signal is proportional to the  difference  between data and MC signal, i.e.\ variable $z_{ij}$, where a proportionality coefficient depends on the muon scaling factor $R_{\upmu,ij}$.

Equation \eqref{eqSmu} can also be rewritten as
\begin{equation}
R_{\upmu,ij} =
  1 + \frac{z_{ij}}{g_{\upmu,i}(\theta) \, S_{1000,ij}^\text{MC}} =
  1 + \frac{S_{1000,j}^\text{mock} - S_{1000,ij}^\text{MC}}{g_{\upmu,i}(\theta) \, S_{1000,ij}^\text{MC}}.
\label{eqRmu}
\end{equation}
Using the mock data set previously built and the QGSJetII-04 MC simulations, the $z_{ij}$ distributions are obtained and shown in Fig.~\ref{fig:44}. 

The total average muon signal of the mock data set $\avg{S_{\upmu,\text{tot}}^\text{mock}}$ can be expressed as
\begin{equation}
\avg{S_{\upmu,\text{tot}}^\text{mock}} =
  \frac{1}{N_\text{tot}} \sum_i \sum^{N_i}_j S_{\upmu,ij}^\text{mock},
\label{eqSum}
\end{equation}
by summing over all primaries $i \in \{\text{p,He,N,Fe}\}$ and where $N_\text{p}=10$, $N_\text{He}=26$, $N_\text{N}=31$, and $N_\text{Fe}=1$ are the number of proton, helium, nitrogen, and iron events, respectively, that have been used to create the mock data set, and $N_\text{tot}=N_\text{p}+N_\text{He}+N_\text{N}+N_\text{Fe}$. Equation \eqref{eqSum} can be expressed as
\begin{equation}
\avg{S_{\upmu,\text{tot}}^\text{mock}} =
  \frac{1}{N_\text{tot}} \sum_i N_i \avg{S_{\upmu,i}^\text{mock}},
\label{eqSum2}
\end{equation}
where $\avg{S_{\upmu,i}^\text{mock}}=\frac{1}{N_i}\sum_j^{N_i}S_{\upmu,ij}^\text{mock}$ is the average over $N_i$ events.
Since $N_i/N_\text{tot}$ is simply the fraction $f_i$, $\avg{S_{\upmu,\text{tot}}^\text{mock}}$ can be rewritten as
\begin{equation}
\avg{S_{\upmu,\text{tot}}^\text{mock}} =
  \sum_i f_i \, \avg{S_{\upmu,i}^\text{mock}}.
\label{eqSum3}
\end{equation}
Using the values of $\avg{S_{\upmu,i}^\text{mock}}$ given in Section \ref{sec1}, we obtain $\avg{S_{\upmu,\text{tot}}^\text{mock}}=(19.78\pm0.22)$\,VEM. We can then rescale the MC data set to retrieve $\avg{S_{\upmu,\text{tot}}^\text{mock}}$,
\begin{equation}
\avg{S_{\upmu,\text{tot}}^\text{mock}} \equiv
  \avg{R_{\upmu,i}} \, \avg{S_{\upmu,i}^\text{MC}}.
\label{eqRet}
\end{equation}
where $\avg{S_{\upmu,i}^\text{MC}}=\frac{1}{N_i}\sum_j^{N_i}S_{\upmu,ij}^\text{MC}$ and $\avg{R_{\upmu,i}}=\frac{1}{N_i}\sum_j^{N_i}R_{\upmu,ij}$ are the averages over $N_i$ events.
In other words, the average total muon signal in mock data corresponds to the average muon signal obtained from MC simulations for a given primary, multiplied by the average muon scaling factor obtained for that primary. The values obtained for the right-hand side of Eq.~\eqref{eqRet} are also reported in Table~\ref{tab2}, along with a summary of the results presented until now. The accuracy with which this equivalence is obtained can be calculated through the ratio
\begin{equation}
k \equiv
  \frac{\avg{R_{\upmu,i}} \, \avg{S^\text{MC}_{\upmu,i}} -
        \sum_i f_i \, \avg{S_{\upmu,i}^\text{mock}}}
       {\sum_i f_i \, \avg{S_{\upmu,i}^\text{mock}}}.
\end{equation}
Values of $k$ are reported in the last column of Table~\ref{tab2}. This method allows us to recover the average muon signal of the mock dataset within ${\sim}9\%$.  The results shown in Table~\ref{tab2} are also a validation of the MC simulation for each primary, as we can recover the total muon signal for each primary.

\begin{table*}
\centering
\begin{tabular}{lcccc}
\br
$i$ & $\avg{R_{\upmu,i}}$ & $\avg{S_{\upmu,i}^\text{MC}}$ & $\avg{R_{\upmu,i}}\,\avg{S_{\upmu,i}^\text{MC}}$ & $k$
\\
\mr
p & $1.35 \pm 0.02$ & $15.57 \pm 0.17$ & $21.02 \pm 0.54$ & $6\%$\\     
He & $1.24 \pm 0.01$ & $17.25 \pm 0.19$ &$21.39 \pm 0.41$ & $8\%$\\
N & $1.11 \pm 0.01$ & $19.37 \pm 0.20$ & $21.50 \pm 0.41$ & $9\%$\\
Fe & $1.00 \pm 0.01$ &$21.62 \pm 0.23$ &  $21.62 \pm 0.44$ & $9\%$\\
\br
\end{tabular}
\caption{Mean value of the muon rescaling parameters $R_{\upmu,i}$ for different primaries $i$. Also, the corresponding mean values of the total muon signal $S_\upmu^\text{MC}$ from QGSJetII-4 model (MC), reconstructed muon signal at 1000\,m expected in the mock data set $\avg{R_{\upmu,i}}\,\avg{S^\text{MC}_{\upmu,i}}$ and the ratio $k$ are listed. The errors shown in the fourth column are the maximum error calculated from $\avg{R_{\upmu,i}} \, \delta S_\upmu^\text{MC} + \avg{S_\upmu^\text{MC}} \, \delta R_{\upmu,i}$, where $\delta R_{\upmu,i}$ and $\delta S_\upmu^\text{MC}$ are the errors listed in the second and third column, respectively.}
\label{tab2}
\end{table*}

\section{Calculation of the $\boldsymbol{\beta}$ exponent}
\label{sec-beta}

The number of muons in an air shower is a powerful tracer of the mass of the primary particle. Simulations  and measurements have confirmed that the number $N_\upmu$ of muons produced rises almost linearly with the primary energy $E$, and increases with a small power of the cosmic-ray mass $A$. This behavior can be understood in terms of the Heitler-Matthews model of hadronic air showers~\cite{heitler}, which predicts $N_\upmu^A=A(E/A\epsilon^\uppi_\text{c})^\beta=N_\upmu^\text{p}\,A^{1-\beta}$, with $\beta\simeq0.92$~\footnote{The $N_\upmu^\text{p}$ is the number of muons for proton shower and $\epsilon^\uppi_\text{c}$ is the critical energy  at which pions decay into muons.}. Detailed simulations of $\beta$ show multiple dependencies on hadronic-interaction properties, like  multiplicity, charge ratio and  baryon anti-baryon pair production~\cite{ulrich}. Thus, measurements of the $\beta$ exponent can effectively constrain the parameters governing hadronic  interactions and improve the accuracy of hadronic models. Assuming that the  average reconstructed muon signal $\avg{S_{\upmu,i}^\text{rec}}$ (see below) is proportional to $ N_\upmu$ and calculating the average logarithm of the muon number $N_{\upmu,i}$ for primary $i$ and iron ($A=56$), we get the expression of $\beta$ given by
\begin{equation}
\beta_i =
  1 -
  \frac{\ln\avg{S^\text{rec}_{\upmu,\text{Fe}}} -
        \ln\avg{S^\text{rec}_{\upmu,i}}}
       {\ln A_\text{Fe} - \ln A_i},
\label{eqBeta}
\end{equation}
where $A_i$ is the mass number of all considered primaries $i$ (except iron).

However, the $\beta$ exponent can also be calculated using the reconstructed muon signal for each primary $i$, e.g.\ $S^\text{rec}_{\upmu,i} \equiv r_{\upmu,i} \, \avg{S^\text{MC}_{\upmu,i}}$. Ideally, we should have $S^\text{rec}_{\upmu,i}=S^\text{mock}_{\upmu,i}$. \emph{Here, the definition of the rescaling factor $r_{\upmu,i}$ is slightly different from the one discussed in the previous section as it rather corresponds to the weight needed to be applied to the MC muon signal of each primary in order to recover the muon signal from the mock data set.} Here $S^\text{rec}_{\upmu,i}$ is  by definition the contribution of the signal for each  primary $i$ to the total muon signal. In this case, the exponent $\beta_i$ can be given by
\begin{equation}
\beta_i =
  1 -
  \frac{\ln(r_{\upmu,\text{Fe}} \, \avg{S^\text{MC}_{\upmu,\text{Fe}}}) -
        \ln(r_{\upmu,i} \, \avg{S^\text{MC}_{\upmu,i}})}
       {\ln A_\text{Fe} - \ln A_i}.
\end{equation}

\begin{figure}
  \centering
  \includegraphics[width=0.51\linewidth]{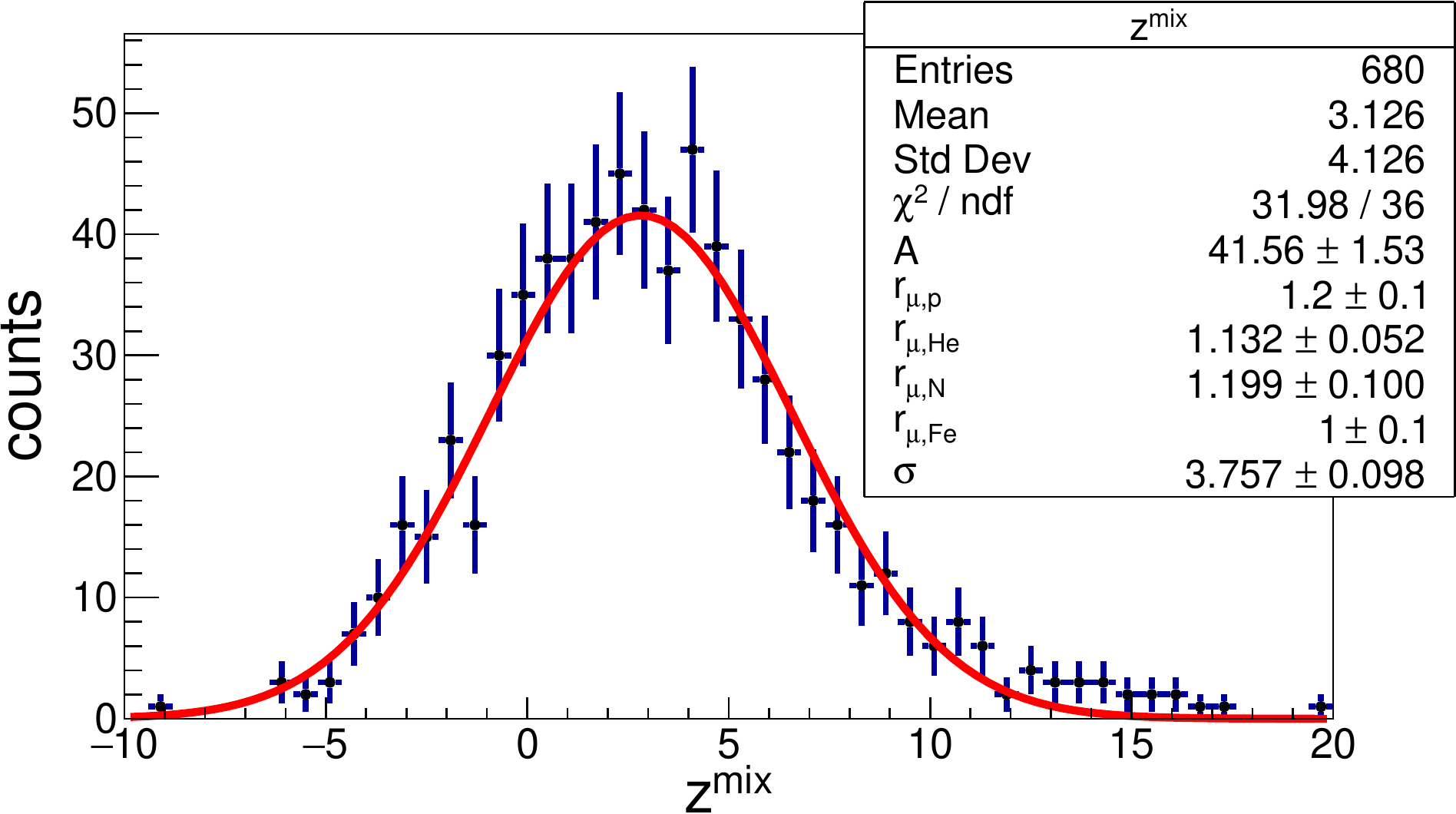}\hfill
  \includegraphics[width=0.46\linewidth]{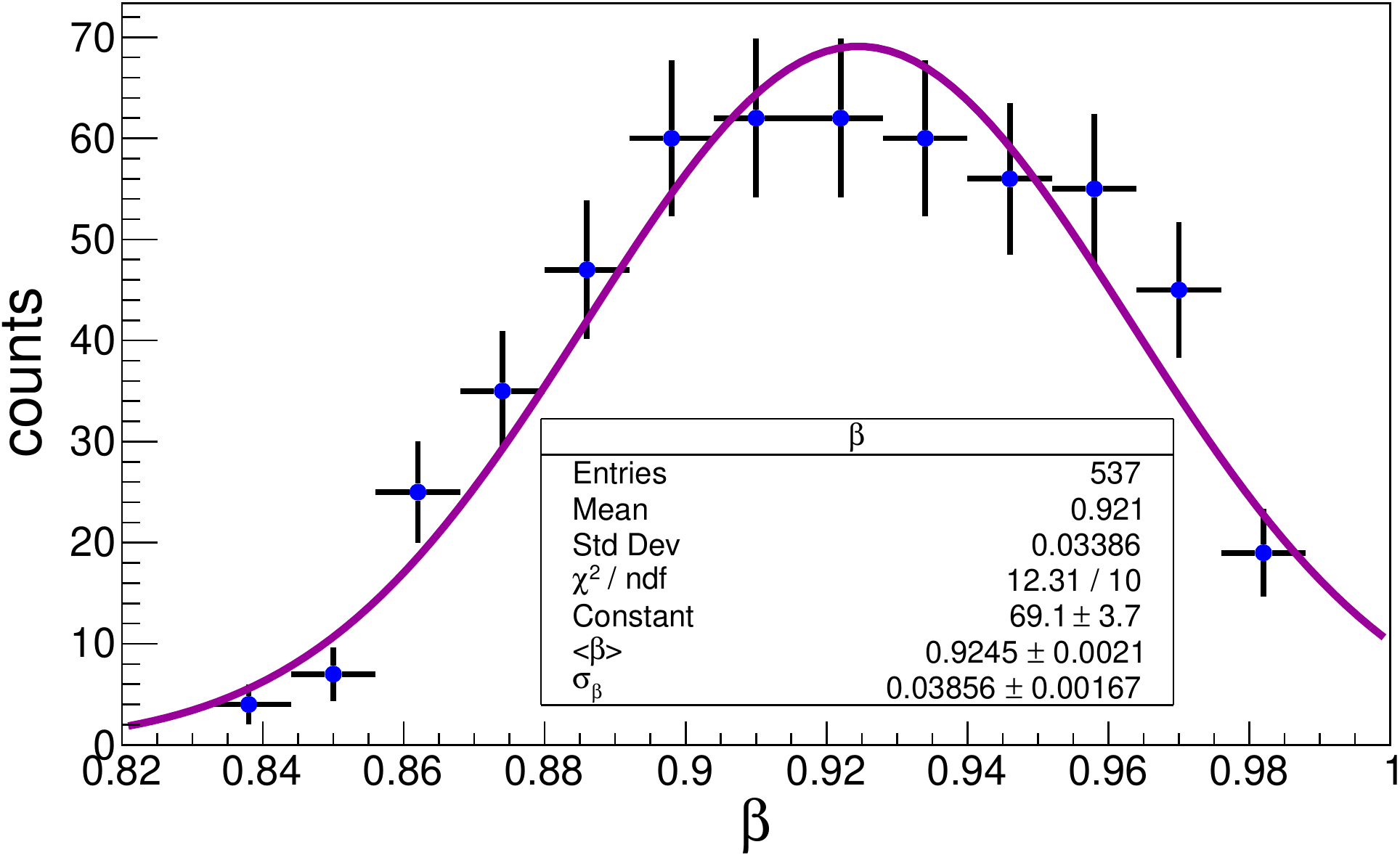}
  \caption{{\bf Left:} $z^\text{mix}$ distribution as described by Eq.~\eqref{eqZmix2} with $f_\text{p}=0.15$, $f_\text{He}=0.38$, $f_\text{N}=0.46$, and $f_\text{Fe}=0.01$. The distribution is fitted with the function described by Eq.~\eqref{eqFitFunc}, with an example of a possible set of fitting parameters $\{A,\sigma,\mathbf{r}_\upmu\}$. \textbf{Right:} Distribution of the average $\beta$ parameter as described by Eq.~\eqref{eqBeta}. The histogram is fitted with a Gaussian function with a mean $\avg{\beta}$ and a standard deviation $\sigma_\beta$ represented by the solid purple line.}
\label{fig:fit1}
\end{figure}

In the following, we show how to compute the $\beta_i$ exponent for a set of hybrid events that consists of certain fractions of events with different primaries. The total signal for the mock and the MC datasets can be expressed as
\begin{align}
\avg{S_{1000}^\text{mock}} &=
  \sum_i f_i \, \avg{S_{1000,i}^\text{mock}} =
  \sum_i f_i (\avg{S_{\text{EM},i}^\text{mock}} +
              \avg{S_{\upmu,i}^\text{mock}}),
\label{eqSum4}
\\
\avg{S_{1000}^\text{MC}} &=
  \sum_i f_i \, \avg{S_{1000,i}^\text{MC}} =
  \sum_i f_i (\avg{S_{\text{EM},i}^\text{MC}} +
              \avg{S_{\upmu,i}^\text{MC}}).
\label{eqSum5}
\end{align}
Again, assuming that in TD-simulations, the electromagnetic component is correctly reproduced, i.e.\ the scaling factor for electromagnetic part is $R_{\text{EM},i}=1$, we can define the overall $z^\text{mix}$ variable as
\begin{align}
\avg{z^\text{mix}} &=
  \avg{S_{1000}^\text{mock}} - \avg{S_{1000}^\text{MC}} =
  \sum_i f_i (\avg{S_{\upmu,i}^\text{mock}} -
              \avg{S_{\upmu,i}^\text{MC}}),
\label{eqSum6}
\\
\avg{z^\text{mix}} &=
  \sum_i f_i \, \avg{S^\text{MC}_{\upmu,i}} \, (r_{\upmu,i} - 1).
\label{eqZmix}
\end{align}
Therefore, for a single event, we can calculate the $z^{\mathrm{mix}}$ variable defined as
\begin{equation}
z^\text{mix}_j \equiv
  S^\text{mock}_{1000,j} - \sum_i f_i \, S^\text{MC}_{1000,ij}.
\label{eqZmix2}
\end{equation}
We consider the same primary fractions for the MC dataset as for the ones used to generate the mock data set, i.e.\ $f_\text{p}=0.15$, $f_\text{He}=0.38$, $f_\text{N}=0.46$, and $f_\text{Fe}=0.01$. The distribution of $z^\text{mix}$ variable is shown in Fig.~\ref{fig:fit1} (left). The $z^\text{mix}$ histogram can be fitted with a Gaussian function described by
\begin{equation}
P(A,\sigma,\mathbf{r}_\upmu) =
  A \exp\left[-\frac{(z^\text{mix}(\mathbf{r}_\upmu)-\avg{z^\text{mix}(\mathbf{r}_\upmu)})^2}{2\sigma^2}\right],
\label{eqFitFunc}
\end{equation}
where fitting parameters are the amplitude $A$, the standard deviation $\sigma$ and four rescaling parameters $\mathbf{r}_\upmu=\{r_{\upmu,\text{p}},r_{\upmu,\text{He}},r_{\upmu,\text{N}},r_{\upmu,\text{Fe}}\}$. Note that following Eq.~\eqref{eqZmix}, the mean of the total muon signal will be proportional to the $\avg{z^\text{mix}}$, and \emph{the factor $r_{\upmu,i}\,\avg{S^\text{MC}_{\upmu,i}}$ is by definition the contribution of the primary $i$ to the total muon signal}. The CERN ROOT~\cite{root} routine Minuit~\cite{minuit} used to fit the histogram requires all these parameters to have initial values when using a user-defined function with multiple parameters. Multiple fits are therefore performed with $r_{\upmu,i}$ between 1 and 2 with steps of 0.025. In this example, correct fits are selected based on physical conditions such as
\begin{equation}
r_{\upmu,\text{p}} \, \avg{S_{\upmu,\text{p}}^\text{MC}} <
r_{\upmu,\text{He}} \, \avg{S_{\upmu,\text{He}}^\text{MC}} <
r_{\upmu,\text{N}} \, \avg{S_{\upmu,\text{N}}^\text{MC}} <
r_{\upmu,\text{Fe}} \, \avg{S_{\upmu,\text{Fe}}^\text{MC}},
\label{eqCon1}
\end{equation}
which simply underlines the fact that the reconstructed muon signal should be larger as the primary gets heavier, and the linearity condition such that
\begin{align}
\left|
  \frac{\ln(r_{\upmu,\text{p}} \, \avg{S_{\upmu,\text{p}}^\text{MC}}) -
        \ln(r_{\upmu,\text{He}} \, \avg{S_{\upmu,\text{He}}^\text{MC}})}
       {\ln A_\text{p} - \ln A_\text{He})} -
  \frac{\ln(r_{\upmu,\text{He}} \, \avg{S_{\upmu,\text{He}}^\text{MC}}) -
        \ln(r_{\upmu,\text{N}} \, \avg{S_{\upmu,\text{N}}^\text{MC}})}
       {\ln A_\text{He} - \ln A_\text{N}}
\right| &< \epsilon \quad\text{and}
\label{eqLin1}
\\
\left|
  \frac{\ln(r_{\upmu,\text{He}} \, \avg{S_{\upmu,\text{He}}^\text{MC}}) -
        \ln(r_{\upmu,\text{N}} \, \avg{S_{\upmu,\text{N}}^\text{MC}})}
       {\ln A_\text{He} - \ln A_\text{N}} -
  \frac{\ln(r_{\upmu,\text{N}} \, \avg{S_{\upmu,\text{N}}^\text{MC}}) -
        \ln(r_{\upmu,\text{Fe}} \, \avg{S_{\upmu,\text{Fe}}^\text{MC}})}
       {\ln A_\text{N} - \ln A_\text{Fe}}
\right| &< \epsilon,
\label{eqLin2}
\end{align}
where $\epsilon=0.10$ is a tolerance of the non-linearity.

The conditions described by Eqs.~\eqref{eqCon1}, \eqref{eqLin1}, and \eqref{eqLin2} are a consequence of the Heitler-Mathews model, which predicts the linear dependence of the muon signal as a function of the logarithm of the primary mass.
The mean values of  $\avg{r_{\upmu,i}}$ distributions, which are reported in Table~\ref{tab3} fall within the uncertainties of the mean true rescaling value $r_\text{true}^\text{MC}$ calculated in Section~\ref{sec1}, i.e.\ $1.10\pm0.04$. The uncertainties of the mean of proton and iron primaries are suspected to stem from the fact that our mock data set contains small numbers of proton and iron events, therefore increasing the uncertainties on the fitting procedure. All the values are reported in Table~\ref{tab3}. With the method proposed in this note, the reconstructed muon signal is overestimated by less than 6\% compared to the muon signal from the mock data set, for all primaries. Using  values of $\avg{S_{\upmu,i}^\text{rec}}$ given in Table~\ref{tab3}, we obtain the total reconstructed muon signal $\avg{S_{\upmu}^\text{rec}}=(20.84\pm0.24)$\,VEM, which differs by approximately +5\% from MC true one, $\avg{S_{\upmu}^\text{mock}}=(19.78\pm0.22)$\,VEM.

\begin{table*}
\centering
\begin{tabular}{lccccc}
\br
$i$ & $\avg{r_{\upmu,i}}$ & $\avg{S_{\upmu,i}^\text{MC}}/\text{VEM}$ & $\avg{S_{\upmu,i}^\text{rec}}/\text{VEM}$ & $\avg{S_{\upmu,i}^\text{mock}}/\text{VEM}$ & $\delta$
\\
\mr
p & $1.142 \pm 0.004$ & $15.57 \pm 0.17$ & $17.78 \pm 0.25$ & $17.30 \pm 0.25$  & 2.7\% \\     
He & $1.167 \pm 0.001$ & $17.25 \pm 0.19$ & $20.13 \pm 0.24$ & $19.03 \pm 0.30$ & 5.8\% \\
N & $1.153 \pm 0.001$ & $19.37 \pm 0.20$ & $22.33 \pm 0.25$ & $21.12 \pm 0.38$ & 5.7\% \\
Fe & $1.148 \pm 0.004$ & $21.62 \pm 0.23$ & $24.82 \pm 0.35$ & $23.42 \pm 0.25$ & 6.0\% \\
\br
\end{tabular}
\caption{Mean values of the muon rescaling factors obtained with the fitting procedure, and of the MC muon signal, the reconstructed and the mock dataset muon signals, for all primaries considered and with $f_\text{p}=0.15$, $f_\text{He}=0.38$, $f_\text{N}=0.46$, and $f_\text{Fe}=0.01$. The overestimation $\delta=(\avg{S_{\upmu,i}^\text{rec}} - \avg{S_{\upmu,i}^\text{mock}}) / \avg{S_{\upmu,i}^\text{mock}}$ of the reconstructed muon signal compared to the one from the mock data set is also provided.}
\label{tab3}
\end{table*}

For each fit fulfilling the conditions described above, we can calculate the averaged
$\beta=\frac{1}{3}\sum_i\beta_i$. The $\beta$ distribution is shown in Fig.~\ref{fig:fit1} (right). The mean of that distribution, $0.924\pm0.002$, is very close to the EPOS-LHC true value of $0.927 \pm 0.003$ (within ${\sim}1\%$)~\cite{calzon}, therefore supporting the effectiveness of the method to estimate the $\beta$ parameter governing the number of muons in hadronic showers.

\section{Summary and conclusion}

The muon problem  currently is one of the hot topics in  cosmic ray  physics, and  for a few years  some  attempts have been made to solve it, but up to now it has not been explained fully.  This is because of   the inaccessibility of certain phase space regions, which are important for  the typical energies of EAS, to accelerator experiments. Exploiting ultra-high energy cosmic rays data, we reach center-of-mas  energies up to 400\,TeV i.e.\ more than 30 times of those attainable at the Large Hadron Collider (LHC)~\cite{lhc}. Thus, an extrapolation of hadronic interaction properties to higher energies is necessary, contributing to systematic uncertainties of the final results. On the other hand, even in the simple Mathews-Heitler model,  increasing the hadronic energy  fraction of interactions by about 5\% per generation,   can lead to about 30\% change in the number of muons after 6 cascade generations. The formation of a Strange Fireball~\cite{sf}, String Percolation~\cite{sp}, Chiral Symmetry Restoration~\cite{csr}, increasing the inelastic cross section~\cite{cs}, or for instance resorting to Lorentz Invariance Violation~\cite{liv} could also explain the muon excess seen in EAS.

The method described in this paper allows us to recover the average muon signal in a hybrid data set, but also offers the possibility to calculate the muon signals for each primary in the considered sample of hybrid events. We show how to compute the $\beta_i$ exponent for a set of hybrid events that consists of a certain fractions of events with different primaries. By using EPOS-LHC simulation as mock dataset and QGSJetII-04 simulations as MC dataset, we can recover the average muon signal in the mock dataset within ${\sim}9\%$, and within less than ${\sim}6\%$ for individual primaries. The average $\beta$ value calculated from the reconstructed muon signal agrees well with the results shown in~\cite{calzon} for EPOS-LHC and QGSJetII-04. The method can be applied to real events to determine the muon signal for each primary, as well as the scaling factor and the $\beta$ exponent. Thus, measurements of the $\beta$ exponent can effectively constrain the parameters governing hadronic  interactions, improve the accuracy of hadronic models, and also show that  ultra-high energy cosmic rays present a great opportunity to explore particle physics beyond the reach of accelerators.

\section*{Acknowledgments}

The authors are very grateful to the Pierre Auger Collaboration for providing the tools necessary for simulation  for this contribution. The authors would like to thank the colleagues from the Pierre Auger Collaboration for all the fruitful discussions. We thank Armando di Matteo  for reading the manuscript and Ruben Concei\c{c}\~ao for valuable discussions. We want to acknowledge the support in Poland from the National Science Centre grant No.~2016/23/B/ST9/01635, grant No.~2020/39/B/ST9/01398 and from the Ministry of Education and Science grant No.~DIR/WK/2018/11 and grant No.~2022/WK/12.

\end{document}